\documentclass{elsart}

\usepackage{cite}

\usepackage{dcolumn}

\begin{document}

\begin{frontmatter}

\title{Amazing variational approach to chemical reactions}

\author{Francisco M. Fern\'{a}ndez \thanksref{FMF}}

\address{INIFTA (UNLP, CCT La Plata-CONICET), Divisi\'{o}n Qu\'{i}mica Te\'{o}rica,\\
Diag. 113 y 64 (S/N), Sucursal 4, Casilla de Correo 16,\\
1900 La Plata, Argentina}

\thanks[FMF]{e--mail: fernande@quimica.unlp.edu.ar}

\begin{abstract}
In this letter we analyse an amazing variational approach to chemical reactions.
Our results clearly show that the variational expressions are unsuitable for
the analysis of empirical data obtained from chemical reactions.
\end{abstract}

\end{frontmatter}

Some time ago Liu and He\cite{LH04} proposed the application of a variational
approach to chemical reactions. Their method consisted of the transformation
of the differential equation for the reaction rate into a kind of Newton
equation that they easily converted into a variational problem for the
mechanical energy. By means of an appropriate ansatz they derived
approximate expressions for the extent of reaction and half--time.

The purpose of this letter is to investigate if those variational
expressions may be of any utility for a chemist.

Liu and He\cite{LH04} considered the chemical reaction
\begin{equation}
nA\rightarrow C+D  \label{eq:ChemReac}
\end{equation}
If $N_{A}(t)$, $N_{B}(t)$, and $N_{C}(t)$ are the number of molecules of the
species $A$, $B$, and $C$, respectively, at time $t$ then Liu and He\cite{LH04}
assumed that $N_{A}(0)=a$, and $N_{B}(0)=N_{C}(0)=0$. If we call
$x=N_{B}(t)=N_{C}(t)$, then we conclude that $N_{A}(t)=a-nx$, where $x$ is
known as the extent of reaction\cite{AD02}. The unique rate
of reaction can be
defined in terms of the extent of reaction as $v=dx/dt$.
Liu and He\cite{LH04}
further assumed that the rate law is given by
\begin{equation}
\frac{dx}{dt}=k(a-x)^{n}  \label{eq:rate_law}
\end{equation}
At this point we stress the fact that this expression is correct only if the
chemical reaction (\ref{eq:ChemReac}) is elementary, otherwise the rate law
may be more complicated. Most chemical reactions are not elementary and
therefore the reaction order and molecularity do not necessarily agree, as
discussed in any book on physical chemistry\cite{AD02}or chemical
kinetics\cite{B60}. What is more, the order of reaction may not even be a
positive integer\cite{AD02,B60}. This fact is known by any undergraduate student
of chemical kinetics but Liu and He seem to be ignorant of it\cite{LH04}.
For concreteness here we assume that the rate law (\ref{eq:rate_law}) is correct.

Liu and He\cite{LH04} tried and solved the differential equation (\ref{eq:rate_law})
approximately by means of a variational approach based on the so--called
semi--inverse method. It consisted in finding the minimum of the integral
expression
\begin{equation}
J=\frac{1}{2}\int_{0}^{\infty }\left[ \left( \frac{dx}{dt}\right)
^{2}+k^{2}(a-x)^{2n}\right] dt  \label{eq:J}
\end{equation}
by means of the variational ansatz
\begin{equation}
x^{var}=a\left( 1-e^{-\eta t}\right)   \label{eq:x_var}
\end{equation}
where $\eta $ is a variational parameter. Notice that $x^{var}(t)$ satisfies
the boundary conditions at $t=0$ and $t\rightarrow \infty $. They found that
the optimal value of the effective first--order rate constant $\eta $ was
given by
\begin{equation}
\eta =\frac{ka^{n-1}}{\sqrt{n}}  \label{eq:eta}
\end{equation}

Liu and He\cite{LH04} argued that chemists and technologists always want to know the
half--time $t_{1/2}=t(x=a/2)$ (which they decided to call halfway).
According to
equations (\ref{eq:x_var}) and (\ref{eq:eta}) the half--time is given
approximately by
\begin{equation}
t_{1/2}^{var}=-\frac{\sqrt{n}\ln (1/2)}{ka^{n-1}}
\label{eq:halftime_var}
\end{equation}
which they showed to be 98\% accurate for $n=2$.

Any textbook on physical chemistry\cite{AD02} or chemical
kinetics\cite{B60}
shows that the exact solution to equation (\ref{eq:rate_law}) is
\begin{equation}
x^{exact}=a\left\{ 1-\frac{1}{\left[ 1+k(n-1)a^{n-1}t\right] ^{1/(n-1)}}%
\right\} ,\,n\neq 1  \label{eq:x_exact}
\end{equation}
and that the exact half--time is given by
\begin{equation}
t_{1/2}^{exact}=\frac{2^{n-1}-1}{k(n-1)a^{n-1}}  \label{eq:halftime_exact}
\end{equation}
Any undergraduate student of chemical kinetics is able to obtain the exact
solution~(\ref{eq:x_exact}) to the differential equation~(\ref{eq:rate_law}) and,
consequently, the exact half--time~(\ref{eq:halftime_exact}). Therefore, there is
no need for approximate methods for this extremely simple problem. However, we
may assume that Liu and He\cite{LH04} chose this exactly solvable problem as a
benchmark for testing their powerful variational approach. We will therefore
analyse if their approximate result is worth the effort.

It is common practice in chemistry to estimate the half--time from
experimental data in order to determine the order of the reaction. Obviously,
an inaccurate expression would lead to an inexact order of reaction.

The variational half--time (\ref{eq:halftime_var}) is reasonably accurate
for $n=2$ because it is exact for $n=1$. The reason is that the variational
ansatz (\ref{eq:x_var}) is the exact solution for a first--order reaction
when $\eta =k$. Notice that it leads to such a result when $n=1$. Table 1
shows that $t_{1/2}^{var}/t_{1/2}^{exact}$ increasingly deviates from unity
as $n$ increases.

The half--time (or half--life) is a particular case of partial reaction
times. We may, for example, calculate the time $t=t_{1/4}$ that has to
elapse for the number of $A$ molecules to reduce to $a/4$ ($x=3a/4$). It is
not difficult to verify that
\begin{equation}
\frac{t_{1/4}^{exact}}{t_{1/2}^{exact}}=2^{n-1}+1
\label{eq:tcuart/tmed_exact}
\end{equation}
From the experimental measure of $t_{1/2}$ and $t_{1/4}$ chemists are able
to obtain the reaction order $n$. However, if they used the variational
expression (\ref{eq:x_var}) they would obtain
\begin{equation}
\frac{t_{1/4}^{var}}{t_{1/2}^{var}}=2  \label{eq:tcuart/tmed_var}
\end{equation}
that is useless for $n \neq 1$.
According to what we have just said it is not surprising that this ratio is
exact for $n=1$. We clearly appreciate that the variational result does not
provide the information that chemists would like to have because it only
predicts first--order reactions. Most probably Liu and He\cite{LH04}
were ignorant of
this fact when they proposed their remarkable variational method.

From the discussion above we conclude that no chemist will resort to
Liu and He's
variational expressions in the study of chemical reactions. There is no
reason whatsoever for the use of an unreliable approximate expression when
one has simple exact analytical expressions at hand. Besides, we have
clearly proved that the variational expressions are utterly misleading.

In a review on asymptotic methods for strongly nonlinear problems He\cite{H06}
came back to this problem and stated that the exact reaction extent for $n=2$ is
\begin{equation}
x_{He}^{exact}(n=2)=a\left( 1-\frac{1}{1-kat}\right)  \label{eq:x(n=2)_He}
\end{equation}
This result is obviously wrong because it exhibits an unphysical pole
at $t=1/(ka)$. From this incorrect expression He\cite{H06} derived a meaningless
negative half--time
\begin{equation}
t_{1/2}^{He}(n=2)=-\frac{1}{ka}  \label{eq:halftime_He_exact_n=2}
\end{equation}
In order to obtain a reasonable agreement with the variational result (\ref
{eq:halftime_var}) He\cite{H06} carried out the following wrong calculation
\begin{equation}
t_{1/2}^{var}(n=2)=-\frac{\sqrt{2}\ln (1/2)}{ka}=-\frac{0.98}{ka}
\label{eq:halftime_He_var_n=2}
\end{equation}
In this way He\cite{H06} managed to obtain two unphysical negative half--times that
agreed 98\%.

If you think that a paper with such a poor scientific quality is uncommon in the
physics literature I will prove you wrong. There has recently been great interest in
pseudo--scientific nonsensical results as we have discussed in several
communications\cite{F07,F08b,F08,F08c,F08d,F08e,F08f,F09a,AF09}. The journals that
publish such extremely low quality papers do not accept comments or criticisms to them.
I recommend the reader to have a look at an approach to population dynamics that
predicts a negative number of rabbits\cite{F08d}. There must be
considerable gains
in publishing such papers that compensate the obvious loss of credibility.

\begin{table}[H]
\caption{Ratio $t_{1/2}^{var}/t_{1/2}^{exact}$ in terms of the reaction order
$n$}
\label{tab:halftime}
\begin{center}
\begin{tabular}{rl}
\hline
$n$ & $t_{1/2}^{var}/t_{1/2}^{exact}$ \\
\hline
2 & 0.98 \\
3 & 0.80 \\
4 & 0.59 \\
5 & 0.41 \\
6 & 0.27 \\
7 & 0.17 \\
8 & 0.11 \\
9 & 0.07 \\
10 & 0.04 \\
\hline
\end{tabular}
\end{center}
\end{table}
\end{document}